\documentclass[amssymb,aps,preprint,onecolumn,prl,floats,showpacs]{revtex4-1}
\usepackage{graphicx,pstricks,float}
\usepackage{natbib}
\usepackage{fancybox}
\usepackage{epsfig}
\usepackage{graphicx}
\usepackage{tabularx}
\usepackage{inputenc}
\usepackage{amsmath,braket,amsfonts,amsmath,mathtools,mathrsfs}
\usepackage{amssymb}
\usepackage{gensymb}
\usepackage{grffile}
\usepackage{relsize}
\usepackage{array}
\usepackage{url,hyperref}
\usepackage{xcolor}
\usepackage{babel}

\usepackage{calrsfs}
\DeclareMathAlphabet{\pazocal}{OMS}{zplm}{m}{n}

\hypersetup{
	colorlinks=true,
	linkcolor = blue,
	filecolor = blue,      
	urlcolor = blue,
	citecolor = blue,
}

\begin{document}
\title{{Antiferromagnetic Skyrmion Crystals in the Rashba Hund's Model on Triangular Lattice}}

\author{Arnob Mukherjee}
\author{Deepak Singh Kathyat}
\author{Sanjeev Kumar}

\affiliation{ Department of Physical Sciences,
Indian Institute of Science Education and Research (IISER) Mohali, Sector 81, S.A.S. Nagar, Manauli P. O. 140306, India
}
\begin{abstract}
Motivated by the importance of antiferromagnetic skyrmions as building blocks of next-generation data storage and processing devices, we report theoretical and computational analysis of a model for a spin-orbit coupled correlated magnet on a triangular lattice. We find that two distinct antiferromagnetic skyrmion crystal (AF-SkX) states can be stabilized at low temperatures in the presence of external magnetic field. The results are obtained via Monte Carlo simulations on an effective magnetic model derived from the microscopic electronic Hamiltonian consisting of Rashba spin-orbit coupling, as well as the Hund's rule coupling of electrons to large classical spins. The two AF-SkX phases are understood to originate from a classical spin liquid state that exists at low but finite temperatures. These AF-SkX states can be easily distinguished from each other in experiments as they are characterized by peaks at distinct momenta in the spin structure factor which is directly measured in neutron scattering experiments. We also discuss examples of materials where the model as well as the two AF-SkX states can be realized.
\end{abstract}

\maketitle
\section{Introduction}
Strongly correlated electrons residing on geometrically frustrated lattices lead to intriguing ordered as well as disordered phases \cite{cage2003,nakatsuji2005,collins1997,gardner2010,seabra2010}. While such systems are extremely challenging to study, suitably motivated approximate treatments not only lead to predictions of remarkable new phases of electronic matter, but also provide new paradigms for understanding the observed electronic properties of solids. Some examples of such phases are quantum and classical spin liquids \cite{paddison2017,shen2016,itou2010,misguich1999,zhu2018,misguich1998}, non-coplanar, non-collinear magnetic states \cite{martin2008,ishikawa2014,kimura2006,yamamoto2014,seki2010} and partially ordered states \cite{ishizuka2013,singhania2020}. Furthermore, the competition between spin-orbit coupling (SOC) and electronic correlations has emerged as one of the most interesting area of fundamental research in recent years \cite{watanabe2010,meetei2015,farrell2014,banerjee2014,sutter2017,paramekanti2018,riera2013}.
In particular, the possibility to tune Rashba SOC via a suitable material growth and design in terms of thin-film multilayers or interfaces has allowed for a realization of SOC induced effects in real materials. One important consequence, with potential applications in data storage and processing technologies, is the observation of skyrmions, antiskyrmions and antiferromagnetic skyrmions in various metals and insulators \cite{andrikopoulos2016,nayak2017,koshibae2016,barker2016,zhang2016}. Indeed, such topological spin textures are considered as building blocks of information storage in the race-track memory devices \cite{tomasello2017,shen2019,fert2013,nagaosa2013}. Antiferromagnetic skyrmions are considered superior to skyrmions as the former do not exhibit skyrmion Hall effect which affects the device performance in case of skyrmions \cite{reichhardt2015,jiang2017,litzius2017}.

In this report, we present the results of our investigations of a prototype model that combines three of the most interesting aspects of electronic problems, namely, geometrical frustration, strong correlations and Rashba SOC. Our main motivation is to understand the physics of antiferromagnetic skyrmion formation in a microscopic electronic Hamiltonian. Most theoretical investigations of formation of skyrmion-like quasiparticles use suitable spin Hamiltonians as a starting point \cite{yi2009,heinze2011,okubo2012,rosales2015}. Instead, here we begin with a microscopic model with itinerant electrons coupled to localized magnetic moments via Hund's rule coupling in the presence of Rashba SOC. Such a model can be realized in thin films or interfaces of transition metal or heavy fermion compounds \cite{caviglia2010,shalom2010,banerjee2013,hwang2012,miron2010,shimozawa2014,samokhin2004}. We explicitly derive a low-energy magnetic Hamiltonian for the triangular lattice for the half-filled insulating case. Given the complex and competing nature of different terms in the Hamiltonian we investigate the low-temperature phases with varying external field via unbiased Monte Carlo simulation technique. In addition to the expected $120^{\circ}$ state and the single-Q spiral states, we identify three non-trivial magnetic phases in the model: (i) a classical spin liquid (CSL) characterized via a diffuse ring pattern centered at the K points of the first Brillouin zone (BZ), (ii) a $3Q$ AF-SkX1 characterized by hexagonal peak pattern in spin structure factor (SSF), and (iii) a qualitatively distinct $6Q$ AF-SkX2 phase that has never been reported before. The degeneracy-induced CSL state can be understood as the parent of both the AF-SkX phases. Our study reports, that not only the AF-SkX states can be described within a microscopic electronic model, but also two distinct AF-SkX phases exist in the triangular lattice model. These two phases can be easily detected in experiments as they lead to qualitatively different peak structure in the neutron scattering data.

\section{Derivation of the Low Energy Spin Hamiltonian}

\noindent
The Hamiltonian describing Rashba electrons coupled to localized magnetic moments residing on a triangular lattice is given by \cite{wenk2012},

\begin{eqnarray} \label{eq:KLM}
H & = & - t \sum_{i,\gamma,\sigma} (c^\dagger_{i, \sigma} c^{}_{i+\gamma, \sigma} + \text{H.c.}) 
- {\textrm i}\lambda \sum_{i,\gamma, \sigma \sigma'} c_{i\sigma}^{\dagger} [\hat{z}\cdot(\boldsymbol{\tau} \times \hat{\bf{\gamma}})]_{\sigma \sigma'} c_{j\sigma'} \nonumber \\ 
& & - J_\text{H} \sum_{i} {\bf S}_i \cdot {\bf s}_i - h_z \sum_{i} S_i^z.
\end{eqnarray}

\noindent
Operator $c_{i\sigma}^\dagger$ ($c_{i\sigma}$) creates (annihilates) an electron at site $i$ with spin $\sigma \in \{\uparrow, \downarrow \}$. $\boldsymbol{\tau}$ is a vector operator with the three Pauli matrices as components. ${\bf S}_i$ denote the localized spins which we assume to be classical vectors with $|{\bf S}_i| \equiv 1$. $t$, $\lambda$ and $J_H$ denote the strengths of hopping amplitude, Rashba coupling and Hund's rule coupling, in that order. Assuming the lattice constant to be unity, $\hat{\gamma} \in \{ {\bf a}_1, {\bf a}_2, {\bf a}_3 \}$ denotes the basis unit vector of the triangular Bravais lattice with ${\bf a}_1$=(1,0), ${\bf a}_2$=(1/2,$\sqrt{3}/2$) and ${\bf a}_3$=(-1/2,$\sqrt{3}/2$). ${\bf s}_i$ is the electronic spin operator. $h_z$ is the strength of Zeeman field applied along $z$ axis, and $\textrm{i} = \sqrt{-1}$. 

The Hamiltonian specified in Eq. (\ref{eq:KLM}) above can be realized in a variety of magnetic compounds comprising of transition metal or rare-earth ions where more than one bands are partially filled. In addition, the existence of Rashba SOC requires inversion symmetry breaking which can be realized in thin films or at interfaces \cite{caviglia2010,shalom2010,caprara2012,banerjee2013,gopinadhan2015,hwang2012,miron2010,wang2012,pesin2012,nakamura2012,king2012}.
We are interested in a situation where charge degree of freedom is completely frozen due to strong correlations and the low energy physics is described in terms of an effective magnetic Hamiltonian. Such a condition is met in Mott insulators where strong Hubbard term disfavours any transfer of charge. In the Hund's model, a similar scenario is realized for large $J_{H}$ at half filling. For large $J_H$, it is useful to work in a site-dependent spin-quantization basis achieved via local SU(2) rotations, given by,

\begin{center}
	$\begin{bmatrix}
	c_{i\uparrow} \\
	c_{i\downarrow}
	\end{bmatrix} 
	=
	\begin{bmatrix}
	\cos(\frac {\theta_i}{2})   & -\sin(\frac {\theta_i}{2}) e^{-\textrm{i} \phi_i} \\
	\sin(\frac {\theta_i}{2}) e^{\textrm{i} \phi_i} & \cos(\frac {\theta_i}{2}) 
	
	\end{bmatrix}  \begin{bmatrix}
	d_{ip} \\
	d_{ia}
	\end{bmatrix}$.
\end{center}

\noindent
Here, $d_{ip} (d_{ia})$  annihilates an electron at site ${i}$ with spin parallel (anti-parallel) to the localized spin. The polar and azimuthal angle pair \{$\theta_i, \phi_i$\} specifies the orientation in three dimensions of the local moment ${\bf S}_i$. 

\noindent
The transformed Hamiltonian takes the form,

\begin{equation}
 H = \sum_{i, \gamma, \sigma \sigma'} \big( g_{i, \gamma}^{\sigma \sigma'} d_{i, \sigma}^{\dagger} d_{i+\gamma, \sigma'} + \text{H.c.} \big) - \frac{J_\text{H}}{2} \sum_i \big(n_{ip} - n_{ia} \big),
\end{equation}
\noindent
where $\sigma \in \{p,a\}$. The transformation to local basis puts the interaction term in a diagonal form. However, the spin dependence now resides in the hopping parameters. The projected hopping amplitudes $ g_{i,\gamma}^{\sigma \sigma'} = t_{i,\gamma}^{\sigma \sigma'} + \lambda_{i,\gamma}^{\sigma \sigma'} $ have contributions from standard tight-binding hopping integral $t$ and the Rashba SOC $\lambda$. Following the 2nd order perturbation approach applied to an isolated pair of sites, we derive the classical super-exchange (CSE) model for the triangular lattice \cite{Mukherjee2020a}.

\vspace{1cm}

\noindent
The parallel to anti-parallel hopping contributions $g_{i,\gamma}^{pa}$, are given by,
\begin{eqnarray} \label{eq:gpa}
 t_{i,\gamma}^{pa} &=& -t \Big[\sin(\frac {\theta_i}{2})  \cos(\frac {\theta_j}{2}) e^{-\textrm{i} \phi_i} -\cos(\frac {\theta_i}{2}) \sin(\frac {\theta_j}{2}) e^{-\textrm{i} \phi_j} \Big], \nonumber \\
 \lambda_{i,{\bf a}_1}^{pa} &=& \lambda \Big[\cos(\frac {\theta_i}{2}) \cos(\frac {\theta_j}{2}) + \sin(\frac {\theta_i}{2})  \sin(\frac {\theta_j}{2}) e^{-\textrm{i} (\phi_i+\phi_j)}  \Big], \nonumber \\
 \lambda_{i,{\bf a}_2}^{pa} &=& \frac{\lambda}{2} \Big[(1-\textrm{i}\sqrt{3})\cos(\frac {\theta_i}{2}) \cos(\frac {\theta_j}{2}) + (1+\textrm{i}\sqrt{3})\sin(\frac {\theta_i}{2})  \sin(\frac {\theta_j}{2}) e^{-\textrm{i} (\phi_i+\phi_j)}  \Big], \nonumber \\
 \lambda_{i,{\bf a}_3}^{pa} &=& -\frac{\lambda}{2} \Big[(1+\textrm{i}\sqrt{3})\cos(\frac {\theta_i}{2}) \cos(\frac {\theta_j}{2}) + (1-\textrm{i}\sqrt{3})\sin(\frac {\theta_i}{2})  \sin(\frac {\theta_j}{2}) e^{-\textrm{i} (\phi_i+\phi_j)}  \Big].
\end{eqnarray}

\noindent
The anti-parallel to parallel contributions, $g_{ij}^{ap}$, are given by,
\begin{eqnarray} \label{eq:gap}
 t_{i,\gamma}^{ap} &=& -t \Big[\cos(\frac {\theta_i}{2}) \sin(\frac {\theta_j}{2}) e^{\textrm{i} \phi_j} - \sin(\frac {\theta_i}{2})  \cos(\frac {\theta_j}{2}) e^{\textrm{i} \phi_i}  \Big], \nonumber \\
 \lambda_{i,{\bf a}_1}^{ap} &=& -\lambda \Big[\cos(\frac {\theta_i}{2}) \cos(\frac {\theta_j}{2}) + \sin(\frac {\theta_i}{2})  \sin(\frac {\theta_j}{2}) e^{\textrm{i} (\phi_i+\phi_j)}  \Big], \nonumber \\
 \lambda_{i,{\bf a}_2}^{ap} &=& -\frac{\lambda}{2} \Big[(1+\textrm{i}\sqrt{3})\cos(\frac {\theta_i}{2}) \cos(\frac {\theta_j}{2}) + (1-\textrm{i}\sqrt{3})\sin(\frac {\theta_i}{2})  \sin(\frac {\theta_j}{2}) e^{\textrm{i} (\phi_i+\phi_j)}  \Big], \nonumber \\
 \lambda_{i,{\bf a}_3}^{ap} &=& \frac{\lambda}{2} \Big[(1-\textrm{i}\sqrt{3})\cos(\frac {\theta_i}{2}) \cos(\frac {\theta_j}{2}) + (1+\textrm{i}\sqrt{3})\sin(\frac {\theta_i}{2})  \sin(\frac {\theta_j}{2}) e^{\textrm{i} (\phi_i+\phi_j)}  \Big].
\end{eqnarray}
\noindent In the above equations, $j = i+\gamma$. The general expression for the second order perturbative energy correction,

\begin{eqnarray} \label{eq: 2nd order correction}
\Delta E^{(2)}_\mathbf{\{S\}} = \mathlarger{\mathlarger{\sum}}_k \frac{ |\braket{\psi_k|H^{\prime}|\psi_0}|^2 }{ E_0-E_k } = -\Bigg[ \frac{|g_{ij}^{pa}|^2}{J_\text{H}} + \frac{|g_{ij}^{ap}|^2}{J_\text{H}} \Bigg],
\end{eqnarray}

\noindent
involves modulus squares of the hopping amplitudes which are obtained from Eq. (\ref{eq:gpa}) and Eq. (\ref{eq:gap}) in the following form:

\begin{eqnarray} \label{eq:SH}
|g^{pa}_{i,{\bf a}_1}|^2 = |g^{ap}_{i,{\bf a}_1}|^2 &=&   \bigg[ \frac{t^2}{2}(1-{\bf S}_i \cdot {\bf S}_j) + \frac{\lambda^2}{2}(1+{\bf S}_i \cdot {\bf S}_j-2S_i^{y}S_j^{y}) - t\lambda(S_i^zS_j^x - S_i^xS_j^z) \bigg], \nonumber \\
|g^{pa}_{i,{\bf a}_2}|^2 = |g^{ap}_{i,{\bf a}_2}|^2 &=&  \bigg[\frac{t^2}{2}(1-{\bf S}_i \cdot {\bf S}_j) + \frac{\lambda^2}{4}\{  2(1+S_i^zS_j^z) - S_i^xS_j^x + S_i^yS_j^y + \sqrt{3}S_i^yS_j^x + \sqrt{3}S_i^xS_j^y \} \nonumber \\ 
& & + \frac{t\lambda}{2} \{ (S_i^xS_j^z - S_i^zS_j^x) + \sqrt{3}(S_i^yS_j^z - S_i^zS_j^y) \} \bigg], \nonumber \\
|g^{pa}_{i,{\bf a}_3}|^2 = |g^{ap}_{i,{\bf a}_3}|^2 &=&  \bigg[t^2(1-{\bf S}_i \cdot {\bf S}_j) + \frac{\lambda^2}{2}\{  2(1+S_i^zS_j^z) - S_i^xS_j^x + S_i^yS_j^y - \sqrt{3}S_i^yS_j^x - \sqrt{3}S_i^xS_j^y \} \nonumber \\ 
& & - \frac{t\lambda}{2} \{ (S_i^xS_j^z - S_i^zS_j^x) - \sqrt{3}(S_i^yS_j^z - S_i^zS_j^y) \} \bigg].
\end{eqnarray}

\noindent
Substituting the expression from Eq. (\ref{eq:SH}) into Eq. (\ref{eq: 2nd order correction}) and taking the sum over all nn pairs, we arrive at the classical super-exchange (CSE) Hamiltonian on a triangular lattice,

\begin{eqnarray}
H_{\textrm{CSE}} &=& -1/J_{\textrm H} \sum_{i,\gamma} \big[ t^2(1-{\bf S}_i \cdot {\bf S}_j) + 2t\lambda  \hat{\gamma'} \cdot ({\bf S}_i \times{\bf S}_j)  \nonumber  \\
& & + \lambda^2(1+{\bf S}_i \cdot {\bf S}_j-2 (\hat{\gamma'} \cdot {\bf S}_i)(\hat{\gamma'} \cdot {\bf S}_j)) \big] - h_z \sum_{i} S_i^z,
\label{eq:ESH}
\end{eqnarray}

\noindent
with $\hat{\gamma'} = \hat{z} \times \hat{\gamma}$. We also note that the Hamiltonian is written in a general form which is valid for any Bravais lattice. 

\begin{figure}[H]
\centering
\includegraphics[width=0.8 \columnwidth,angle=0,clip=true]{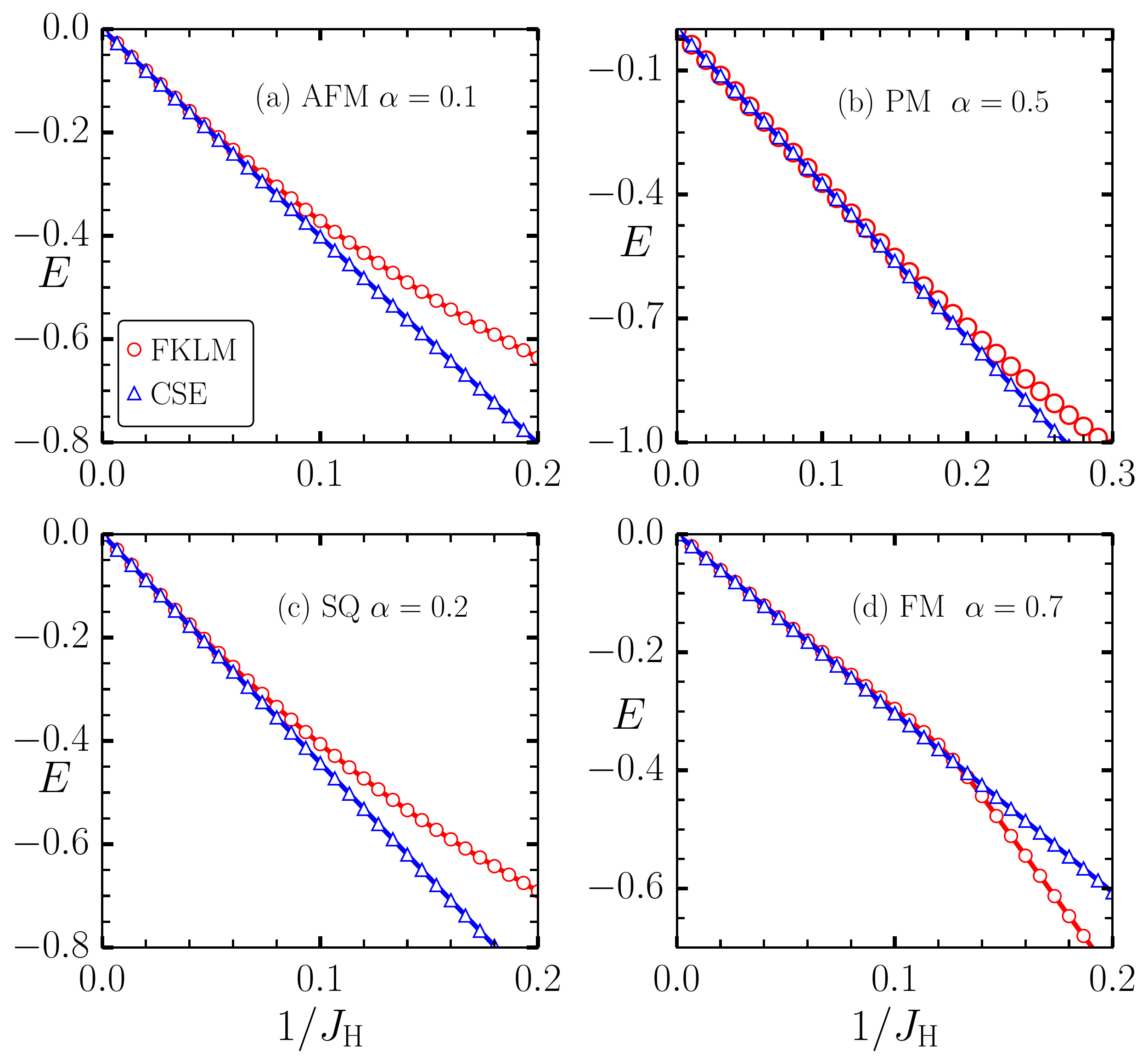}
\caption{ $1/J_H$ dependence of energy per site calculated via exact diagonalization of the Rashba Hund's Hamilonian (red circles) and that calculated from the low-energy spin Hamilonian (blue triangles) for different magnetic states and for different $\alpha$: (a) 120$^{\circ}$ state, (b) random configuration representative of a paramagnetic state, (c) single-Q spiral state, and (d) ferromagnetic state. 
}
\label{e-compare}
\end{figure}

We parameterize by $\alpha$ the relative strength of the Rashba coupling as compared to hopping parameter as $t = (1-\alpha) t_0$ and $\lambda = \alpha t_0$, where $t_0=1$ sets the reference energy scale. The resulting model consists of antiferromagnetic coupling terms along with anisotropic terms resembling Dzyaloshinskii-Moriya (DM) and Kitaev-like interactions \cite{Kathyat2020a,kathyat2021a,Mukherjee2020a} . The ground state of the Hamiltonian Eq. (\ref{eq:ESH}) for $\alpha=0$ is the well known three-sublattice $120^{\degree}$ state which stabilizes due to geometrical frustration. In absence of external magnetic field ($h_z = 0$), increasing $\alpha$ favours non-collinear spin arrangement due to the presence of DM terms. However, the frustrating nature of the DM and Kiteav-like terms leads to a large degeneracy of states, as discussed by us for the case of square lattice \cite{Mukherjee2020a}. One consequence of this large degeneracy is the presence of entropically stabilized filamentary domain states at low temperatures.
Before proceeding with the investigations of the magnetic properties of the electronic Hamiltonian Eq. (\ref{eq:KLM}) in terms of the effective low energy CSE Hamiltonian Eq. (\ref{eq:ESH}), we explicitly check the validity of CSE Hamiltonian by comparing energies of different magnetic states obtained within the exact and approximate Hamiltonians. The energies calculated via the CSE Hamiltonian match very well with the exact values, provided $1/J_H < 0.1$ (see Fig. \ref{e-compare}). Note that the large $J_H$ expansion is only valid when the parallel and antiparallel bands are split and the chemical potential resides in the gap. For the triangular lattice tight-binding bands, such splitting will occur for $J_H = 9$ for the largest bandwidth corresponding to ferromagnetic background.
Therefore, the comparison in Fig. \ref{e-compare} shows that the effective Hamilonian is quantitatively accurate as long as the gap-opening condition is satisfied.

\section{Results and Discussion}
\subsection{A. Phases in the absence of external magnetic field}

\begin{figure}[H]
\centering
\includegraphics[width=0.98 \columnwidth,angle=0,clip=true]{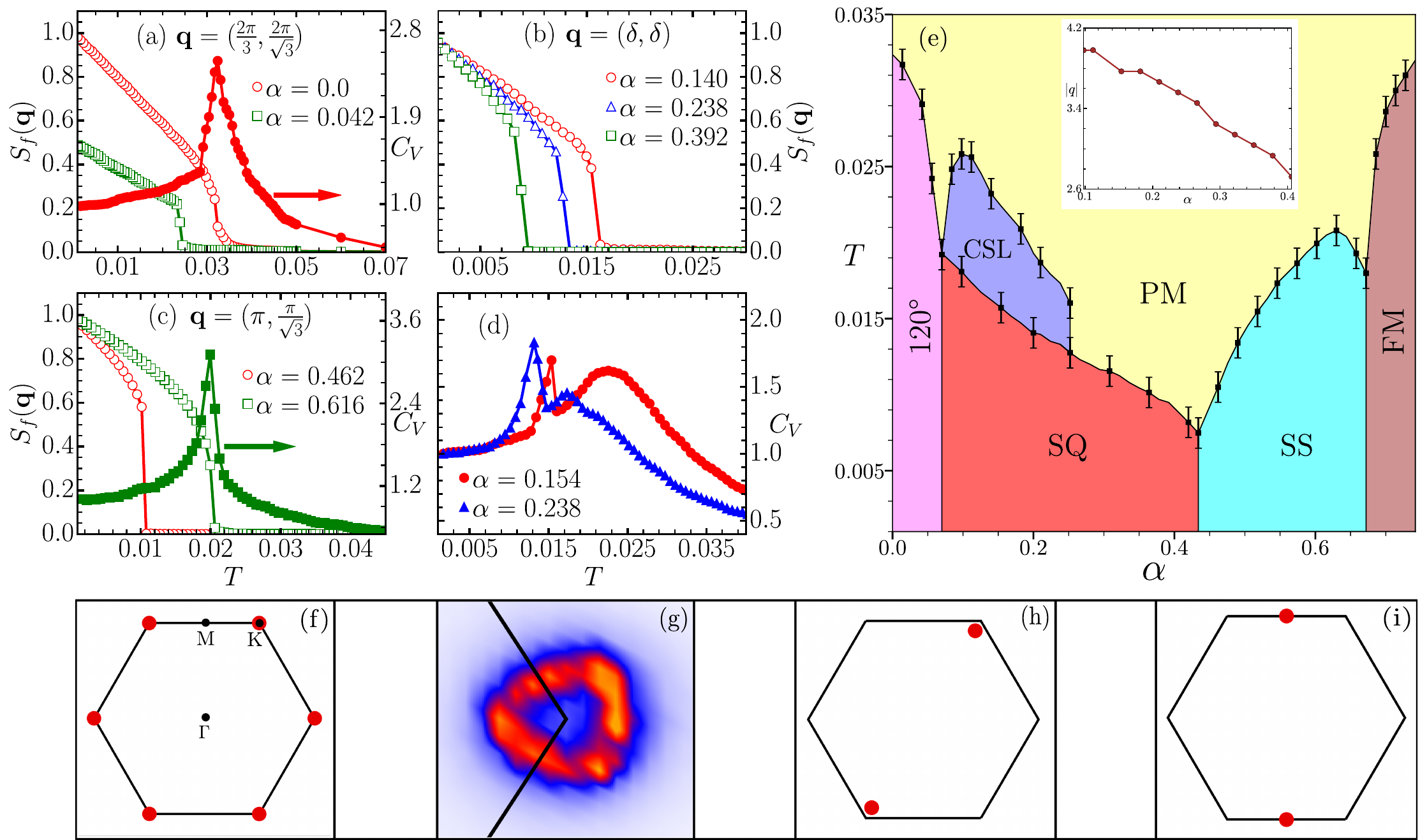}
\caption{ (a)-(c) Temperature dependence of spin structure factor (open symbols) at different values of ${\bf q}$, and for different values of relative Rashba coupling strength $\alpha$. The right y-axis in (a) and (c) is for specific heat (filled symbols) corresponding to one of the $\alpha$ values. (d) Specific heat as a function of temperature showing a broad hump followed by a sharp peak for two values of $\alpha$. (e) Phase diagram obtained 
by tracking the features in the SSF and the specific heat with the different phases described as follows: the $120^{\circ}$ denotes the well known three-sublattice order on triangular lattice. Single-Q (SQ) denotes a state where SSF displays peak only at a pair ${\pm \bf q}$ of momentum. Spin stripe (SS) phase consists of ferromagnetic lines oriented antiferromagnetically w.r.t. neighboring lines. Classical spin liquid (CSL) denotes a phase with short-range correlations but no long-range ordering. (f)-(i) SSF plots for the four non-trivial phases displayed in the phase diagram. Note that the CSL state is characterized by a diffuse ring-like pattern centered at the K point.
A zoomed in view is shown in panel (g). Inset in (e) shows variation in the magnitude $q$ of the relevant wave-vector $\bf{q}$ with $\alpha$ for SQ state.
}
\label{PD1}
\end{figure}

\noindent
In this section, we present the Monte Carlo simulation results for the CSE Hamiltonian Eq. (\ref{eq:ESH}) (see Methods). We begin by studying the model in the absence of external magnetic field. In the limit $\alpha \rightarrow 0$, the Hamilonian reduces to a simple antiferromagnetic Heisenberg model and the lowest energy is obtained for a three-sublattice $120^{\degree}$ arrangement of spins. We track the temperature dependence of the spin structure factor (SSF), as obtained in simulations, at relevant value of ${\bf q}$. The $120^{\degree}$ state is characterized by a SSF peak at ${\bf q}_0 = (2\pi/3, 2\pi/\sqrt{3})$, and the symmetry related points (see Fig. \ref{PD1}(f)). For small values of $\alpha$, the SSF at ${\bf q}_0$ exhibits an order parameter like rise upon lowering temperature (see Fig. \ref{PD1}(a)). The sharp upturn point is identified as the ordering temperature, which is further verified via specific heat calculations (filled symbols in Fig. \ref{PD1}(a)). Beyond a critical value of $\alpha$, the $120^{\degree}$ ground state is destabilized in favour of a single-Q (SQ) spiral state with SSF peak located at $\pm{\bf q}$ for one specific ${\bf q}$ (see Fig. \ref{PD1}(h)). In contrast to the $120^{\degree}$ state, the SQ state lifts the three-fold degeneracy as one pair of K points is spontaneously selected from three equivalent choices. The ordering temperature for the SQ state, as inferred from the temperature dependence of SSF at relevant ${\bf q}$, monotonically decreases upon increasing $\alpha$ (see Fig. \ref{PD1}(b)). For $\alpha > 0.43$, a specific SQ state which consists of ferromagnetic (FM) chains oriented antiferromagnetically, labelled as spin stripe (SS) state, is stable over a wide range of $\alpha$. This is characterized by a peak in the SSF at a pair of M points (see Fig. \ref{PD1}(i)). Similar to the SQ state, the SS state is also three-fold degenerate and the degeneracy is spontaneously lifted. Furthermore, the specific heat displays a sharp peak at the temperature corresponding to the on-set of SSF peak (see Fig. \ref{PD1}(c)) allowing us to reliably infer the ordering temperatures.
Eventually a FM state, characterized by magnetization, becomes the ground state in the limit of strong Rashba coupling. As discussed above, the transitions from the high temperature paramagnetic (PM) state to any of the ordered states can be described with the help of the relevant peak in the SSF. These transitions are also identified as sharp peaks in the specific heat as shown in Fig.\ref{PD1}(a),(c)). We find that for most of the $\alpha$ values, the sharp rise in the order parameter is accompanied by a peak in the specific heat. However, for $0.07 < \alpha < 0.25$ we find a broad hump feature in the specific heat in addition to a sharp peak (see Fig. \ref{PD1}(d)). On a careful observation of the SSF, we find the presence of diffuse circular patterns centered about the K-points of the first BZ (see Fig. \ref{PD1}(g)). This allows us to identify this finite-temperature phase as a classical spin liquid, similar to the one reported in the square lattice \cite{Mukherjee2020a}. We summarize our results of Monte Carlo simulations in the absence of magnetic field as a $T-\alpha$ phase diagram (see Fig. \ref{PD1}(e)). While there is a similarity with the square lattice phase diagram, it is surprizing to note that the geometrical frustration inherent in the triangular geometry disfavors the zero-field skyrmion state found in the square lattice \cite{Mukherjee2020a}. The zero-field skyrmion state reported in the square-lattice model consists of skyrmions packed in a square geometry which is not compatible in a triangular lattice. Therefore, the SS state is preferred over the zero-field skyrmion crystal state for $0.43 < \alpha < 0.67$.

\subsection{B. Phases in the presence of external magnetic field}
\begin{figure}[H]
\centering
\includegraphics[width=0.96 \columnwidth,angle=0,clip=true]{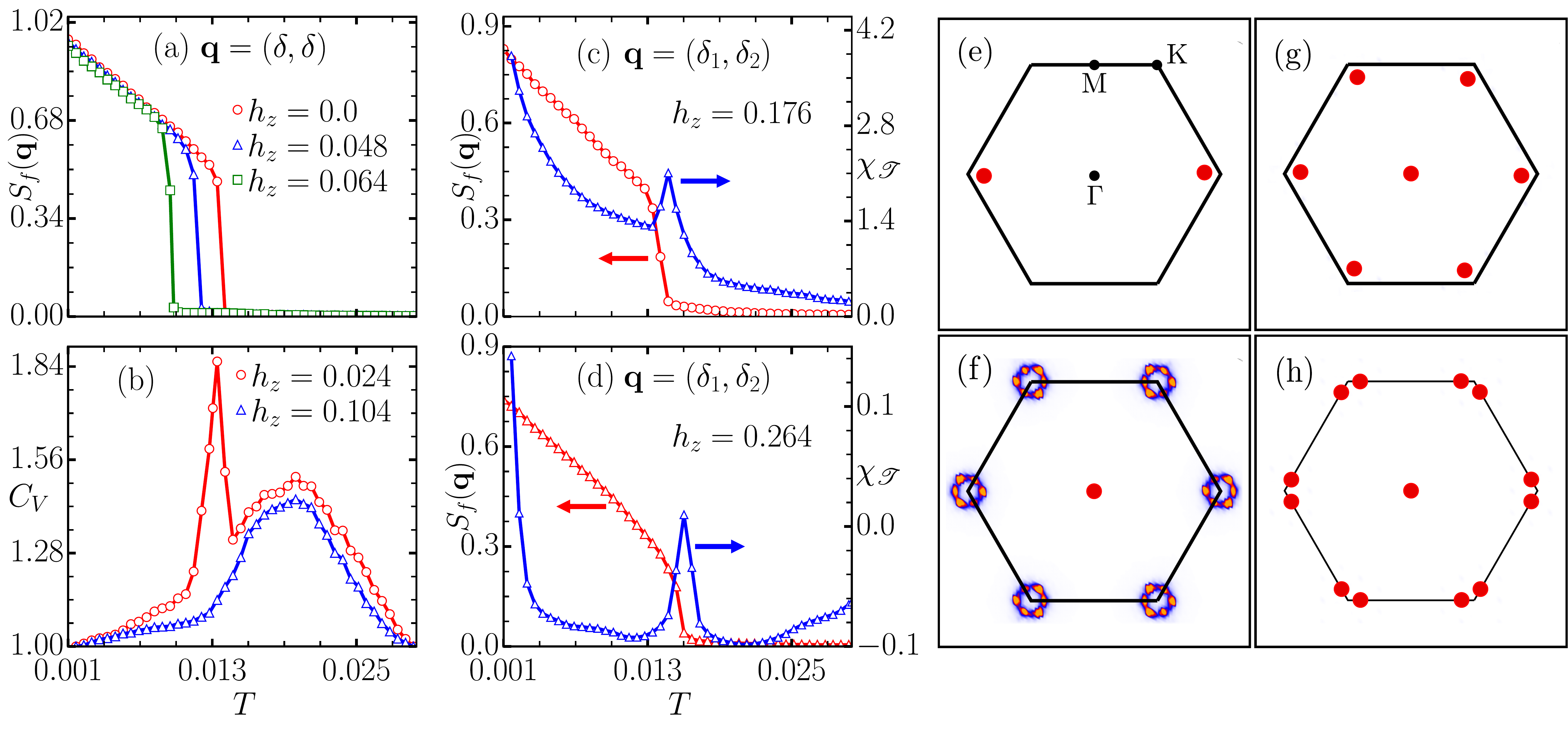}
\caption{(a) SSF peak as a function of temperature for SQ state, (b) specific heat variation with temperature to identify CSL state transition, (c)-(d) variation of SSF peak (left axis) and topological susceptibility (right axis) for AF-SkX1 and AF-SkX2 states respectively. (e)-(h) SSF peak locations for four magnetic states displayed in the phase diagram (Fig. \ref{pd}) SQ, CSL, AF-SkX1 and AF-SkX2 respectively. 
}
\label{op-vs-T}
\end{figure}

Having established the zero-field phase diagram for the triangular lattice, we now discuss the effect of Zeeman field on the magnetic states.
In Fig. \ref{op-vs-T} (a)-(d), we show temperature dependence of SSF, $C_V$ and topological susceptibility (see Methods) at finite $h_z$.
We discuss the case of $\alpha = 0.2$ as a representative of finite Rashba SOC.
For small $h_z$, the ground state remains a three-fold degenerate SQ spiral state discussed in the previous subsection (Fig. \ref{op-vs-T} (e) displays another choice of the degenerate ${\bf q}$ point). The ordering temperature, as inferred from the SSF at the relevant ${\bf q}$ (see Fig. \ref{op-vs-T}(a)),  decreases with increasing $h_z$. Interestingly, over a moderate range of $h_z$ values ($0.07 < h_z < 0.14$) the SSF does not display any prominant peak. Specific heat also shows only a broad hump and no sharp anomaly (see Fig. \ref{op-vs-T}(b)). We confirm the existence of diffuse circular pattern in SSF at low temperatures in this range of $h_z$, similar to that shown in Fig. \ref{PD1}(g) (see Fig. \ref{op-vs-T}(f)). Therefore, we conclude that the CSL state gets stabilized down to very low temperatures for finite Zeeman field. This stabilization of a short-range ordered spin-liquid state by application of external magnetic field is unusual and is analogous to melting of a solid under pressure as magnetic field in spin systems is analogous to external pressure in real solids. Upon increasing $h_z$ further, we find 
two exotic ordered states: a 3$Q$ antiferromagnetic skyrmion crystal, henceforth labelled as AF-SkX1, (see Fig. \ref{op-vs-T}(g)) and a novel antiferromagnetic skyrmion crystal, henceforth labelled as AF-SkX2, with SSF peaks on the boundaries (straight line joining nearest pairs of K and M points) of the first BZ (see Fig. \ref{op-vs-T}(h)) \cite{rosales2015,hayami2019} .

\noindent
We further characterize the two multi-$Q$ states with the help of skyrmion density ($\mathcal{T}$) and topological susceptibility ($\chi_{\mathcal{T}}$) (see Methods). Indeed, the topological susceptibility peaks at the on-set temperature of the multi-$Q$ order as inferred from the SSF (see Fig. \ref{op-vs-T}(c),(d)) in both the skyrmion states, AF-SkX1 and AF-SkX2. The peaks in $\chi_{\mathcal{T}}$ are clear indications of non-topological to topological phase transitions.

\begin{figure}[H]
\centering
\includegraphics[width=0.98 \columnwidth,angle=0,clip=true]{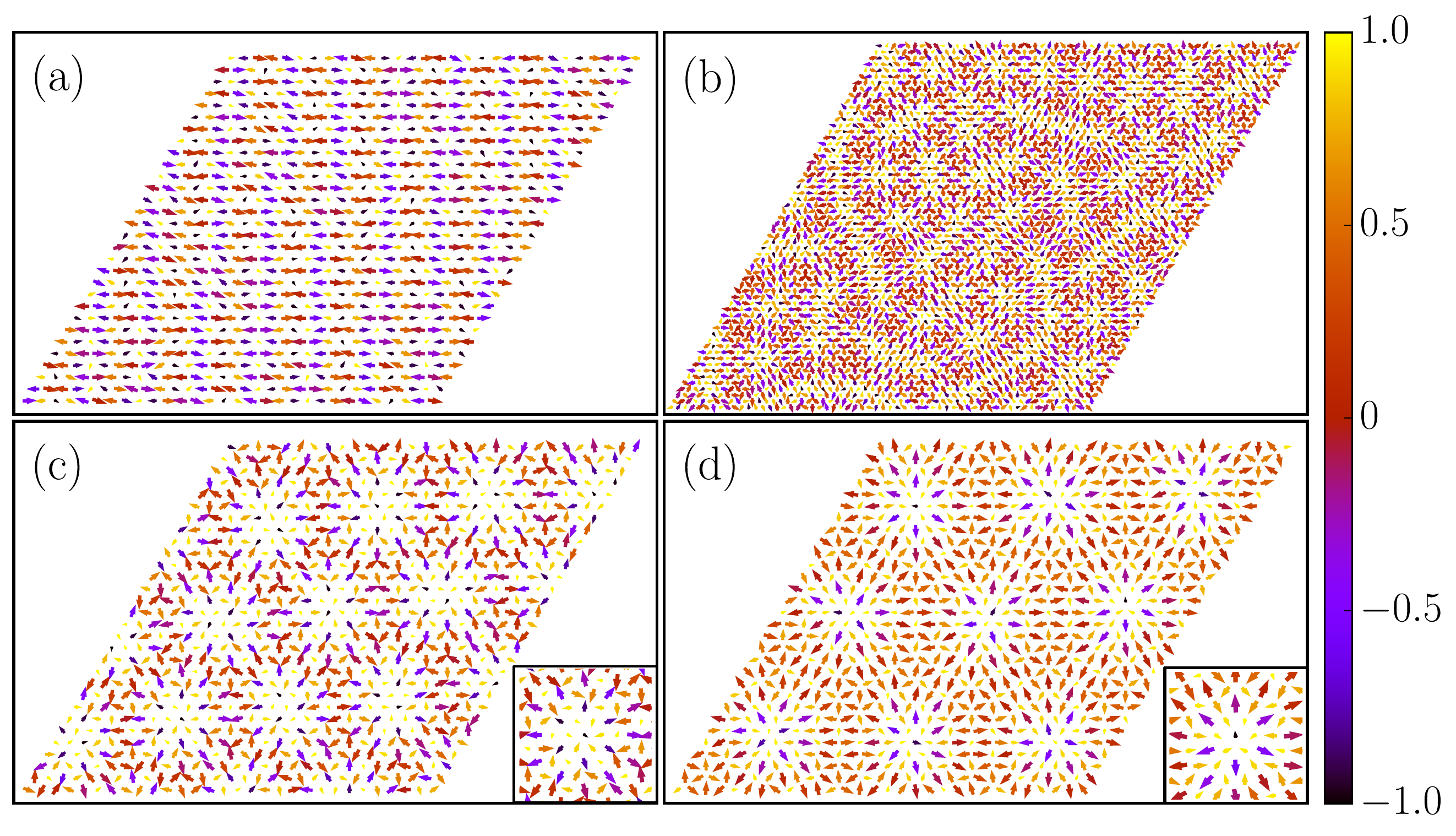}
\caption{Low-temperature spin configurations for (a) single-Q spiral, (b) classical spin liquid, (c) AF-SkX1 and (d) AF-SkX2 states. The $z$-component is represented by the color bar and the planar components by the arrow lengths and directions. For clarity, we only display a $30 \times 30$ section of the simulated lattice in panels (a), (c), (d) and $60 \times 60$ for (b). The insets in (c) and (d) show the zoomed-in view of spins in the core of skyrmions.
}
\label{sf-spin}
\end{figure}

\noindent
In Fig. \ref{sf-spin} we show the evolution with increasing Zeeman field of low temperature magnetic states via representative spin configurations. The SQ state consists of spins spiraling in the $xz$ plane with the ordering wavevector residing on the $x$ axis in the reciprocal space (the corresponding SSF is shown in Fig. \ref{op-vs-T}(e)). Upon increasing magnetic field the system enters a short-range ordered CSL phase. A typical spin configuration in the CSL state consists of filamentary domain segments (see Fig. \ref{sf-spin}(b)). The existence of such a disordered state relies on the presence of an unusual degeneracy of the SQ spirals that involved a simultaneous change of the spiral wavevector and the spin plane. This is discussed by us in recent papers \cite{Kathyat2020a,Mukherjee2020a}. Indeed, the CSL state is similar to the antiferromagnetic string state discussed in \cite{Mukherjee2020a}, with the difference that the CSL state emerges in the background of  $120^{\degree}$ state on triangular lattice. It is interesting to note that some of the filaments existing in the CSL state are short, and hence acquire a skyrmion-like modulations of the spins. This is suggestive that the CSL state is unstable towards a state hosting skyrmions.

Indeed, this is confirmed as increasing magnetic field leads to the formation of AF-SkX1 state. A typical configuration of spins in the AF-SkX1 state is shown in 
Fig. \ref{sf-spin}(c) where a triangular arrangement of antiferromagnetic skyrmions is observed in the background of $120^{\degree}$ state. With a further increase in the strength of Zeeman field, we find the AF-SkX2 as the ground state (see Fig. \ref{sf-spin}(d)). While it is difficult to distinguish between AF-SkX1 and AF-SkX2 looking at the real-space spin configurations, the SSF for AF-SkX2 is qualitatively different from that for AF-SkX1 (compare Fig. \ref{op-vs-T}(g) and \ref{op-vs-T} (h)). This can be interpreted as a superposition of two counter-rotated triangular arrangements of the skyrmions. Another interpretation is that the AF-SkX1 state is closer to SQ (SSF peaks on the $\Gamma$-K line) while the AF-SkX2 is closer to SS (SSF peaks on the K-M line). Therefore, AF-SkX1 and AF-SkX2 can be visualized as originating from the underlying classical spin liquid state with circular pattern in SSF (see Fig. \ref{sf-spin}(f)) by lifting the degeneracy in two different ways.

\begin{figure}[H]
\centering
\includegraphics[width=0.98 \columnwidth,angle=0,clip=true]{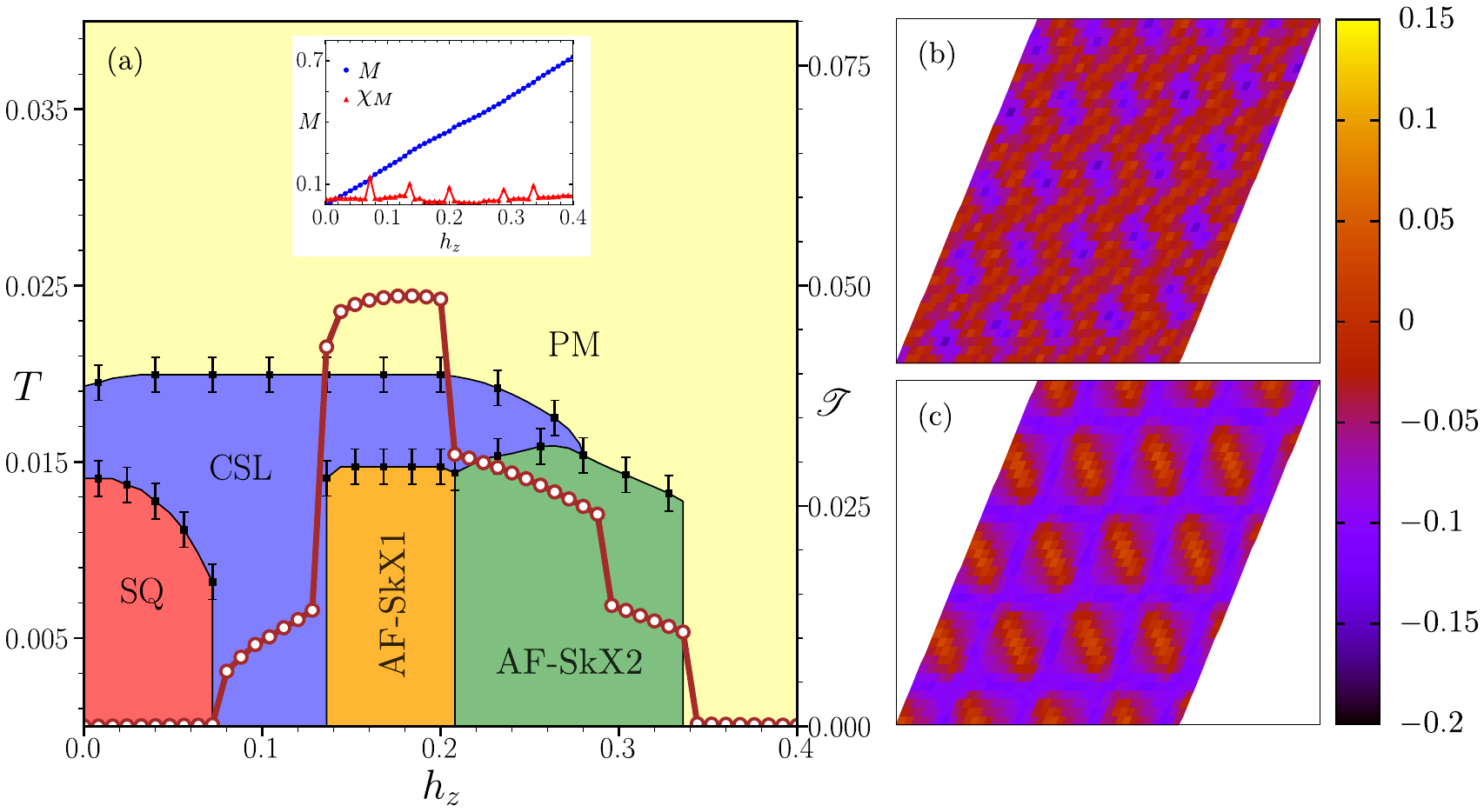}
\caption{(a) The temperature versus magnetic field phase diagram of the Hamilonian Eq. \ref{eq:ESH}. The different phase boundaries are inferred from the order parameter plots shown in Fig. \ref{op-vs-T}. The right axis shows the variation of skyrmion density $\chi_{\mathcal{T}}$ with external magnetic field $h_z$.
Field dependence of magnetization (blue) and that of its derivative (red) are shown in inset. The real-space maps of the skyrmion density for the two antiferromagnetic skyrmion phases (b) AF-SkX1, (c) AF-SkX2.
}
\label{pd}
\end{figure}

\noindent
Our main findings are summarized in the form of $T$ vs. $h_z$ phase diagram shown in Fig. \ref{pd}(a). We discover three non-trivial states in our study. A liquid-like short range ordered state existing between the PM and the SQ state at zero magnetic field becomes stable at low temperatures with increasing magnetic field.  
The AF-SkX1 becomes the ground state near $h_z = 0.14$ which then destabilizes in favour of AF-SkX2 near $h_z = 0.21$.
The boundaries seperating different phases were inferred from a combination of temperature dependence of relevant components of SSF, specific heat and topological susceptibility as discussed before. The open circles display the variation of skyrmion density ($\mathcal{T}$) as a function of applied field at low temperatures. Note that the presence of phase boundaries is clearly reflected in the sharp changes in the skyrmion density.
The existence of a finite skyrmion density in the CSL state indicates the existence of a isolated skyrmions in this phase when the filaments lengths become of the same order as their width (see \ref{sf-spin}(b)). The sharp jump within AF-SkX2 state does not represent a phase transition as the SSF remains qualitatively identical on two sides of the discontinuity. Inset in Fig. \ref{pd}(a) shows magnetization, $M_z$, (blue) and magnetic susceptibility, $\chi_M$, (red) as a function of applied field. Note that the phase changes affect the manner in which magnetization increases with applied field and this gets clearly reflected in the peak structure in $\chi_M$ that exactly matches the indicative phase transitions shown by the skyrmion density variation.

We observe that the total skyrmion density remains almost unchanged in the $h_z$ window corresponding to the AF-SkX1 state. This suggests that the AF-SkX1 state is highly incompressible, and is similar to the packed skyrmion phase discussed by us in a recent paper \cite{kathyat2021a}. In contrast, the skyrmion density in the AF-SkX2 state gradually decreases upon increasing magnetic field. The step wise reduction of $\chi_{\mathcal{T}}$ in AF-SkX2 is a finite size effect, which can accommodate only particular number of skyrmions with the imposed periodic conditions. In continuum limit it is expected that $\chi_{\mathcal{T}}$ should smoothly go to zero.

\section{Conclusion}
\noindent
Starting from a microscopic Rashba-Hund's model on a triangular lattice, we derived an effective magnetic model in the insulating limit. A comprehensive Monte Carlo simulation study of the model uncovers a variety of intriguing magnetic phases. In particular, we find two distinct antiferromagnetic skyrmion crystals as the ground states of the model in the presence of external magnetic field. The effective magnetic model allows us to understand the origin of these two AF-SkX states. Existence of a short-range ordered state characterized by circular diffuse pattern in the SSF serves as the parent of the two SkX states. These two states correspond to two different ways of breaking the degeneracy present in the classical spin liquid. A realization of the model studied here can be achieved in the interfaces and heterostructures of transition-metal oxides (TMOs) along (111) direction. The transition-metal (TM) ions with large moment, such as Mn, Fe etc., are particularly relevant for validating our approximation of a classical spin. The neccessary features to realize our model can be found in many real materials \cite{geselbracht1990,wang2015,luo2011,li2014}. One candidate material is GdI$_2$, in which Gd ions in the $4f^75d^1$ state form a triangular lattice arrangement and electrons from partially filled $d$ bands are coupled to localized $f$ bands \cite{taraphder2008,kasten1984}. In a recent study by Chakhalian et al., the complex oxides A$_2$B$_2$O$_7$ in [111] directional growth opens up a new route where triangular arrangement of high atomic number transition metal ions induce strong spin-orbit coupling \cite{chakhalian2020}. Other potential candidate materials are LaFeO$_3$, LaMnO$_3$, LaFeO$_3$/LaCrO$_3$ \cite{zhu2011}, (LaMnO$_3$)$_2$/(LaScO$_3$)$_4$ \cite{he2012,weng2015} bilayers [111]. In LaMnO$_3$ half-filled t$_{2g}$ electrons and e$_g$ are coupled via Hund's coupling and the bilayers of LaScO$_3$ provide significant spin-orbit coupling.
Given that the low-energy magnetic Hamilonian for a Rashba coupled Mott insulator will have the identical form to what we derived here for the Hund's model, our results regarding the existence of AF-SkX states are also relevant to Mott-insulators on triangular lattices. Since these two skyrmion states can be easily distinguished based on the spin structure factor, our results provide a clear prediction for their observation in neutron scattering experiments.

\section{Methods}
\noindent
We simulate the spin Hamiltonian Eq. (\ref{eq:ESH}) via the Classical Monte Carlo technique based on conventional heat bath method \cite{binder1993monte}. Periodic boundary conditions are implemented along each direction. Temperature parameter is reduced in small steps starting at high temperature to capture the phase transition from paramagnetic to ordered state. For a given value of $T$ and $h_z$, single spin updates are performed by proposing a new spin configuration from a set of uniformly distributed points on the surface of a unit sphere. The new configuration is accepted based on the standard Metropolis algorithm \cite{metropolis1953equation,hastings1970}. A Monte Carlo run at each magnetic field and temperature consist of $\sim 1 \times 10^5$ Monte Carlo steps (MCSs) for equilibration and twice the number for calculations of the desired observables. For detailed exploration of parameter space we used lattice size $N = 60 \times 60$, and the stability of results is ensured by simulating sizes up to $N = 120 \times 120$ for some selected parameter values. For simulations in the presence of external magnetic field, we use the field cooled protocol, where the temperature is lowered in the presence of finite external field.

\noindent
The various phases, obtained via Monte Carlo simulations, can be distinguished from the corresponding real-space spin textures (see Fig. \ref{sf-spin}). Additionally, we have calculated various physical observables to precisely identify the phase transitions. We calculate the magnetization ($M$), magnetic susceptibility ($\chi_M$), specific heat ($C_V$) and the topological susceptibility ($\chi_\mathcal{T}$) \cite{amoroso2020}, defined as,

\vspace{-1.2cm}

\hspace{-1.8cm}
\begin{minipage}{0.5\textwidth}
\begin{equation}\label{eA1}
\begin{split}
M                &= \frac{1}{N} \bigg \langle \sum_i S_i^z \bigg \rangle, \\
\chi_M           &= \frac{dM}{dh_z}, \\
C_V              &= \frac{d \langle E \rangle}{dT}, \\
\chi_\mathcal{T} &= \frac{\mathcal{T}^2 - \mathcal{T}^2}{NT}
\end{split}
\end{equation}
\end{minipage}
\hspace{1.5cm}
\begin{minipage}{0.5\textwidth}
\vspace{1.8cm}
\begin{figure}[H]
\centering
\includegraphics[width=0.4 \linewidth,keepaspectratio=true]{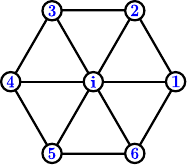}
\caption{Schematic diagram showing locations of nn sites of a central site in the triangular lattice.
}
\label{tri-latt}
\end{figure}
\null
\par\xdef\tpd{\the\prevdepth}
\end{minipage}

\vspace{-0.3cm}

\noindent
The angular brackets denote the Monte-Carlo average of the quantity, $ \langle E \rangle = \frac{1}{N} \langle H_{CSE} \rangle $, and $\mathcal{T}$ denotes the discretized skyrmion density, given as \cite{rosales2015},
\begin{equation}
 \centering
 \mathcal{T} = \frac{1}{4\pi} \Bigg \langle  \sum_i A_i^{(12)} \text{sgn} [\mathcal{L}^{(12)}_{i}] + A_i^{(45)} \text{sgn} [\mathcal{L}^{(45)}_{i}]  \Bigg \rangle,
\end{equation}

\begin{equation}
 \centering
 \mathcal{L} = \frac{1}{8\pi} \Bigg \langle  \sum_i \mathcal{L}_i^{(12)} + \mathcal{L}_i^{(45)} \Bigg \rangle,
\end{equation}

\noindent
where, $A_i^{(ab)} = ||({\bf S}_{i_a} - {\bf S}_{i}) \times ({\bf S}_{i_b} - {\bf S}_{i})||/2$ is the local area of the surface spanned by three spins on every elementary triangular plaquette ${\bf r}_i,{\bf r}_a,{\bf r}_b$. Here $\mathcal{L}_{i}^{(ab)} = {\bf S}_i.({\bf S}_{i_a} \times {\bf S}_{i_b})$ is the so-called local chirality and ${\bf r}_i, {\bf r}_1 - {\bf r}_5$ (see Fig. \ref{tri-latt}) are the sites involved in the calculation of $\mathcal{T}$.

\noindent
Most importantly, we also compute the component resolved spin structure factor (SSF) to characterize the conventional ordered magnetic phases. The SSF is given by,

\begin{eqnarray}
 S_f({\bf q}) &=& S^{x}_f({\bf q}) + S^{y}_f({\bf q}) + S^{z}_f({\bf q}), \nonumber \\
 S^{\mu}_{f}({\bf q}) &=& \frac{1}{N^2} \bigg \langle \sum_{ij} S^{\mu}_i S^{\mu}_j~ e^{-{\rm i}{\bf q} \cdot ({\bf r}_i - {\bf r}_j)} \bigg \rangle.
\label{SSF}
\end{eqnarray}

\noindent
with $\mu = x,y,z$.

\section{Acknowledgment}
We thank Sudhanshu Shekhar Chaurasia for valuable discussions and technical assistance. We acknowledge the use of the computing facility at IISER Mohali.

\bibliographystyle{apsrev4-1} 

%

\end{document}